\documentstyle[preprint,aps,floats,epsfig]{revtex}

\def\book#1[[#2]]{{\it#1\/} (#2).}
\def\am#1 #2 #3.{{\it Ann.\ Math.\ \bf#1} #2 (#3).}
\def\apj#1 #2 #3.{{\it Astrophys.\ J.\ \bf#1} #2 (#3).}
\def\atmp#1 #2 #3.{{\it Adv.\ Theor.\ Math.\ Phys.\ \bf#1} #2 (#3).}
\def\cmp#1 #2 #3.{{\it Commun.\ Math.\ Phys.\ \bf#1} #2 (#3).}
\def\comnpp#1 #2 #3.{{\it Comm.\ Nucl.\ Part.\ Phys.\  \bf#1} #2 (#3).}
\def\cqg#1 #2 #3.{{\it Class.\ Quant.\ Grav.\ \bf#1} #2 (#3).}
\def\epl#1 #2 #3.{{\it Europhys.\ Lett.\ \bf#1} #2 (#3).}
\def\grg#1 #2 #3.{{\it Gen.\ Rel.\ Grav.\ \bf#1} #2 (#3).}
\def\jmp#1 #2 #3.{{\it J.\ Math.\ Phys.\ \bf#1} #2 (#3).}
\def\ijmpd#1 #2 #3.{{\it Int.\ J.\ Mod.\ Phys.\ \bf D#1} #2 (#3).}
\def\mpla#1 #2 #3.{{\it Mod.\ Phys.\ Lett.\ \rm A\bf#1} #2 (#3).}
\def\ncim#1 #2 #3.{{\it Nuovo Cim.\ \bf#1\/} #2 (#3).}
\def\npb#1 #2 #3.{{\it Nucl.\ Phys.\ \rm B\bf#1} #2 (#3).}
\def\phrep#1 #2 #3.{{\it Phys.\ Rep.\ \bf#1\/} #2 (#3).}
\def\pla#1 #2 #3.{{\it Phys.\ Lett.\ \bf#1\/}A #2 (#3).}
\def\plb#1 #2 #3.{{\it Phys.\ Lett.\ \bf#1\/}B #2 (#3).}
\def\pr#1 #2 #3.{{\it Phys.\ Rev.\ \bf#1} #2 (#3).}
\def\prd#1 #2 #3.{{\it Phys.\ Rev.\ \rm D\bf#1} #2 (#3).}
\def\prl#1 #2 #3.{{\it Phys.\ Rev.\ Lett.\ \bf#1} #2 (#3).}
\def\prs#1 #2 #3.{{\it Proc.\ Roy.\ Soc.\ Lond.\ A.\ \bf#1} #2 (#3).}

\def\half{\textstyle{1\over2}}
\def\third{\textstyle{1\over3}}

\def\sech{\,{\rm sech}}

\newcommand{\be}{\begin{equation}}
\newcommand{\ee}{\end{equation}}
\newcommand{\bea}{\begin{eqnarray}}
\newcommand{\eea}{\end{eqnarray}}
\newcommand{\bml}{\begin{mathletters}}
\newcommand{\eml}{\end{mathletters}}

\begin{document}
\preprint{DTP/00/107, EHU-FT/0011, hep-th/0012100}
\draft
\tighten

\title{Black Holes on Thick Branes}
\author{Roberto Emparan$^1$, Ruth Gregory$^2$,
Caroline Santos$^{2,3}$.}
\address{~$^1$ Departamento de F{\'\i}sica Te\'orica,
Universidad del Pa{\'\i}s Vasco, \\ Apdo.\ 644, E-48080 Bilbao, Spain\\
$^2$ Centre for Particle Theory, 
Durham University, South Road, Durham, DH1 3LE, U.K.\\
$^3$ Departamento de F\'\i sica da Faculdade de Ci\^encias da Universidade do
Porto,Rua do Campo Alegre 687, 4150-Porto, Portugal.}
\date{\today}
\setlength{\footnotesep}{0.5\footnotesep}
\maketitle

\begin{abstract}
The interplay between topological defects (branes) and black holes has
been a subject of recent study, motivated in part by interest
in brane-world scenarios. In this paper we analyze in detail the
description of a black hole bound to a domain wall (a two-brane in four
dimensions), for which an exact description in the limit of zero wall
thickness has been given recently. We show how to smooth this singular
solution with a thick domain wall. We also show that charged extremal
black holes of a size (roughly) smaller than the brane thickness expel
the wall, thereby extending the phenomenon of flux expulsion. Finally,
we analyze the process of black hole nucleation {\it on} a domain wall,
and argue that it is preferred over a previously studied mechanism of
black hole nucleation {\it away} from the wall.
\end{abstract}

\pacs{PACS numbers: 04.40.-b, 04.70.Bw, 04.70.Dy, 
11.27.+d \hfill hep-th/0012100}
\section{Introduction}

The study of the interaction of topologically nontrivial field theoretic
solutions and black holes in four dimensions has yielded some
interesting insights in recent years, revising our understanding of the
classic ``no-hair'' theorems for black holes \cite{NHT}, with the
realisation that black holes can have `dressed' horizons \cite{LNW}, or
even topological hair, in the guise of cosmic strings, extending to
infinity \cite{AGK}. An important feature of these (and other, related,
solutions in the literature) is that they explicitly include
gravitational back reaction of the defect on the black hole spacetime.
For a localised defect, this is clearly a dressed version of a
Reissner-Nordstrom (RN) black hole, however, for an extended defect,
such as the cosmic string, the significant global aspects of the
spacetime solution must be taken into account before claims of black
hole hair can be validated. Fortunately, the spacetime of an
infinitesimally thin string with a black hole is known: the
Aryal-Ford-Vilenkin (AFV) \cite{AFV} solution for a thin string piercing
a black hole (a RN black hole with a wedge cut out), the Israel-Kahn
metric \cite{IK} for two black holes suspended in unstable equilibrium
by two strings ending on the event horizons, and the C-metric \cite{KW}
corresponding to a black hole being accelerated by a string extending to
infinity. All of these metrics can have their conical deficits smoothed
by a realistic vortex core model \cite{AGK,SMTH}. 

Also in recent times, domain walls (and other defects) have become a
subject of intense study from the point of view that our universe might
be a brane, or defect, (see \cite{RSA} for pioneering work) sitting in
some higher dimensional spacetime. The motivation has come partly
because string or M-theory appears to admit a phase in which our world
appears as a `wall' \cite{UDW}, but also because of the exciting
phenomenological possibility of an unusual resolution of the hierarchy
problem \cite{HPR}. Most attention has been focussed on the case where
our universe is a gravitating domain wall \cite{EXO,RS}, however, higher
codimension compactifications have been considered \cite{HCD}. A general
feature of these solutions is that four-dimensional gravity is recovered
on the brane universe \cite{RS,GT}, at least perturbatively, although
the question of nonlinear effects, such as black holes on the brane
\cite{CHR}, remains an interesting open one, necessitating a study of
the problem in one dimension less \cite{EHM}: i.e.\ a black hole living
on a wall in four spacetime dimensions.

A natural question, when considering our universe as a brane, is to
investigate models in which our universe is, quite literally, a defect,
namely, a `soliton' solution to some higher dimensional field theory.
This approach was taken in \cite{HCD}, and for the domain wall in
five dimensional anti-de Sitter space, by Gremm \cite{GRM}. 
It appears that one can smooth out the `singular' wall by modelling 
it with the core of a topological domain wall. In a similar fashion
to the infinitesimally thin brane, one can ask questions about strong
gravity on the thick brane, namely a black hole intersecting a thick brane.
This is the question we are interested in in this paper, and for the
same reason as the infinitesimally thin wall, we will examine this issue in
four-dimensional gravity. 

Recently, it was shown numerically that a topological domain wall
could sit through a Schwarzschild black hole \cite{MYIIN} in the
absence of gravitational back reaction. In the light of the results 
for the vortex solution, \cite{AGK}, this is perhaps not surprising, 
however, the issue of gravitational back reaction is particularly
important in this set-up. The gravitational field of a domain wall
was found some time ago \cite{IS}, and in a coordinate system natural
to the wall (i.e.\ exhibiting planar symmetry) was found to be 
time-dependent -- in stark contrast to the static conical 
metric of the cosmic string -- with a de-Sitter like expansion along
the spatial coordinates of the wall. Later, as the global structure
of the wall spacetime was better understood, it was realised that
the horizon singularities of the wall spacetime were removable by
transforming into a `bulk-based' coordinate system, in which 
spacetime is flat, and consists of the interior of two hyperboloids
in Minkowski spacetime glued together \cite{GWG}. Space is compact
in the domain wall spacetime, and the horizon a consequence of the 
acceleration of the bubble. Placing a black hole on the wall therefore
involves placing a black hole on this accelerating wall with its
compact space. An additional question is how charge on the black hole
affects the domain wall. At first sight, one might think that using a
Reissner-Nordstrom black hole, rather than Schwarzschild, for the 
background field theory solution should make no difference, however, this
ignores the issue of flux expulsion. For an extreme black hole, there
is a phenomenon of flux expulsion \cite{CCES,CEG,BEG} in which if the vortex
is thick enough, the black hole will expel its flux, causing the field to
remain in its symmetric state on the horizon. Such a flux expulsion is not
dependent on the vortex being local, it occurs for global strings \cite{BEG} 
and for pure flux p-branes \cite{CEG}, therefore we might well expect a 
similar phenomenon to occur for the domain wall.  All of this discussion
however, hinges on not only the existence of a suitable thin wall metric with
a black hole, but on this being a thin wall limit of a smooth thick wall
metric with a black hole sitting on it. Note that unlike the vortex, the
strong gravitational effect of the domain wall will mean that as soon as
we include gravitational back reaction, the whole nature and
global structure of the spacetime will change. Indeed, even in the
absence of a black hole, there is a closely related phenomenon of wall
non-formation: if the thickness of the wall is
too great compared to the inverse mass of the scalar field forming it,
then it is not possible to form a domain wall \cite{BV,BCG}.

This paper addresses the issues raised above. Fortunately, the metric of
an infinitesimally thin wall has been found in \cite{EHM} for the case
of a wall in adS spacetime, by using the C-metric for the accelerating
wall, sliced in two and identified. In this paper, we show how to smooth
such singular solutions with a thick domain wall thereby demonstrating
the smooth wall/black hole spacetime. We will work with the solutions
where the cosmological constant in the bulk of spacetime vanishes, but
extrapolation to non zero cosmological constant should be
straightforward. As expected, the space is compact, and the black holes
are accelerated along with the wall. We also explore the question of
`flux' expulsion for the wall, proving analytically that such expulsion
occurs, and deriving bounds for the mass of the black hole (in wall
units) at which it must occur. We then analyze the process of black hole
nucleation on a domain wall. Unlike the decay of strings, \cite{SMTH},
the net result of this process is not the disintegration of the wall,
but rather a pair creation of black holes in the presence of the
accelerating wall.

The layout of the paper is as follows: In the next section we review and
extend the work of \cite{EHM}, deriving an appropriate infinitesimal
wall + black hole metric and analyzing its thermodynamics. In section
\ref{sect:twa} we consider the field equations of the domain wall in the
background of a black hole (both RN and the C-metric), deriving an
analytic ``thin-wall'' approximation which will be useful for the
problem of gravitational back-reaction, and then demonstrating, in
section \ref{sect:expel}, the phenomenon of flux expulsion. Section
\ref{sect:gbr} deals with gravitational back reaction, and section
\ref{sect:nbh} with the nucleation of black holes on the wall. Finally,
we summarize our results in section \ref{sect:concl}, and discuss the
possible consequences and extensions to the scenario with a three-brane
in five dimensions, of relevance to brane world models.

\section{The black hole-wall metric }

\subsection{Constructing the solution}

We start by deriving the equivalent of the AFV solution for the vortex, namely
an infinitesimally thin domain wall with a black hole. We begin with the
C-metric \cite{KW}
\be
ds^2 = \frac{1}{A^2\,\,(x+y)^2}\,\,\left[ F(y) dt^2 - \frac{dy^2}{F(y)} 
-G(x) d\varphi^2 - \frac{dx^2}{G(x)} \right]\, ,
\label{Cmet}
\ee
where $G(x)= 1-x^2 - 2mA x^3 - q^2 A^2 x^4 = -F(-x)$. A bulk
cosmological constant $\Lambda$ could be easily incorporated, and in
particular a negative one may be of interest to discuss toy models for
the Randall-Sundrum scenario (see \cite{EHM}). For simplicity, we set
$\Lambda=0$: the results in sections \ref{sect:twa}, \ref{sect:expel}
and \ref{sect:gbr} can be easily extended to non-zero $\Lambda$. On the
other hand, issues such as global structure, thermodynamics and
instantons as studied here, extend qualitatively to all cases where the
geometry induced on the brane is deSitter, even if $\Lambda$ is positive
or negative.

In general the
quartic $G(\xi)$ will have 4 roots, and we will take the parameters such
that they are all real, and labeled as $\xi_1\leq \xi_2 <
\xi_3<0<\xi_4\leq1$. The coordinates in (\ref{Cmet}) are restricted to
$x\in[\xi_3,\xi_4]$, and $y\in(-x,\infty)$; $y = -\xi_3, -\xi_2, -\xi_1
(>0)$ \ are the acceleration horizon, and outer and inner black hole
horizons, respectively. For $m\neq 0$ there is a singularity at
$y=\infty$, which corresponds
to the central singularity of the black hole. Note also that in general
this spacetime has a conical deficit as $x\to\xi_3,\xi_4$, one of which
(say at $\xi_4$) can be eliminated by setting the periodicity of
$\varphi$ to be $\varphi \in [0,{4\pi\over |G'(\xi_4)|}]$. The remaining
conical deficit at $x=\xi_3$ has the interpretation of a string pulling
the accelerating black hole away to infinity. The gauge potential for
the Maxwell field is ${\bf A} = qy {\bf d}t$ for an electric black hole,
or ${\bf A} = q (x-\xi_4) {\bf d} \varphi$ for a magnetic black hole.

The
conventional definitions of mass (say, ADM) cannot be applied to obtain
the mass of the accelerating black hole\footnote{See, however, below.},
however, for a small black hole, i.e., $m,q\ll 1/A$, the geometry
approaches the Reissner-Nordstrom solution, which allows us to identify
approximately the black hole mass as $m/G$. Indeed, it is
useful to have the approximate values for the roots for a small black
hole,
\bea\label{approot}
\xi_1=-{1\over r_-A}+O(mA),&\quad
&\xi_2=-{1\over r_+A}+O(mA),\nonumber\\
\xi_3=-1-m A+O(m^2A^2),&\quad
&\xi_4=1-m A+O(m^2A^2),
\eea
with $r_\pm=m\pm\sqrt{m^2-q^2}$ (we are taking $m\geq q$; the
extremality bound is precisely $\xi_2\geq \xi_1$). The charge of
the black hole can be measured by integrating the flux on a sphere that
surrounds it (e.g., at $y={\rm const}$), and is given by 
\bea 
Q&=&{1\over
4\pi}\int F_{x\varphi}dxd\varphi={\Delta\varphi\over
4\pi}\left(A_\varphi(x=\xi_4)-A_\varphi(x=\xi_3)\right)\nonumber\\
&=&{1\over qA^2}{1\over (\xi_4-\xi_2)(\xi_4-\xi_1)}=q+O(mA)\, , 
\eea
where $\Delta\varphi=4\pi/|G'(\xi_4)|$ is the period of $\varphi$.

To construct the wall-black hole (WBH) metric we follow the standard
Israel construction, according to which the tension $\sigma$ of a domain
wall is given by the discontinuity of the extrinsic curvature,
$[K_{ij}]=4\pi G\sigma h_{ij}$, with $h_{ij}$ the metric induced on the
wall. Following \cite{EHM}, an appropriate totally umbilic surface
($K_{ij}\propto h_{ij}$) can
be found at
$x=0$. This has normal $ n = {1\over A y} dx$,
induced metric
\be
ds^2 = {1\over A^2y^2} \left [ F(y) dt^2 - {dy^2\over F(y)} -
d\varphi^2 \right ]\, ,
\label{indmet}\ee
and extrinsic curvature 
\be
K_{ij} = A h_{ij}\, .
\label{extc}
\ee

We have chosen the conical singularity to lie at $x=\xi_3$, i.e., on the
side $x<0$ of this surface. As a consequence, if we form the WBH
solution by taking two copies of the side $x>0$ and glue them together
along $x=0$, the string will have disappeared from the spacetime.
The construction is equivalent to substituting $|x|$ for $x$ in (\ref{Cmet}).
Note however, that the gauge potential remains unaltered. This is
evident for the case of an electric potential (which does not depend on
$x$), while for a magnetic potential the change in sign in $x$ is
cancelled by a corresponding reversal of ${\bf d}\varphi$.
The Israel construction implies that the tension of the wall thus
formed is 
\be
\sigma={A\over 2\pi G}\, . 
\ee
On the other hand, the charge of the black hole changes now to
\bea
Q&=&2{\Delta\varphi\over
4\pi}\left(A_\varphi(x=\xi_4)-A_\varphi(x=0)\right)\nonumber\\
&=&{2\over qA^2}{\xi_4\over
(\xi_4-\xi_3)(\xi_4-\xi_2)(\xi_4-\xi_1)}\, ,
\eea
but, to leading order in $mA$, we still have $Q\simeq q$.

The metric induced on the wall takes a particularly interesting form if
we introduce a radial coordinate $r=1/(Ay)$, and $t=A\bar t$. Then,
(\ref{indmet}) becomes 
\be
ds^2=\left(1-{2m\over r}+{q^2\over r^2}-A^2r^2\right)d\bar
t^2-{dr^2\over
1-{2m\over r}+{q^2\over r^2}-A^2r^2}-r^2d\varphi^2
\label{desit}\ee
which is exactly the same as the equatorial section of the
four dimensional Reissner-Nordstrom-de Sitter solution.

\begin{figure}
\begin{center}\leavevmode  %
\epsfxsize=6cm 
\epsfbox{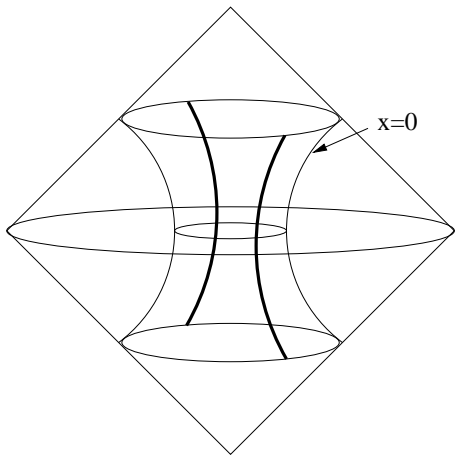}
\end{center}
\caption{Conformal diagram for the embedding of the hyperbolic surface
$x=0$ in Minkowski spacetime (circles in this picture are actually
spheres). The thick lines correspond to $y=+\infty$,
which track the worldlines of the black holes in the C-metric.} 
\label{fig:wallinmink}
\end{figure}

The domain wall that we have constructed actually contains {\it two} black
holes, sitting at antipodal points of a spherical domain wall. In
order to see this, let us have a closer look at the global structure of
the WBH spacetime. It is helpful
to consider first the situation where the black holes are absent from
the wall, i.e.,
$q=m=0$. In that case, the metric is
\be
ds^2={1\over A^2(x+y)^2}\left[(y^2-1)dt^2-{dy^2\over y^2-1}-{dx^2\over
1-x^2}-(1-x^2)d\varphi^2\right]\, .
\label{nobh}\ee
The acceleration horizon is at $y=+1$. If we now
change to coordinates $(T,X,Y,Z)$, using
\bea
X^2-T^2={y^2-1\over A^2(x+y)^2},&&\qquad Y^2+Z^2={1-x^2\over A^2(x+
y)^2}\nonumber\\
{T\over X}=\tanh t,&&\qquad {Z\over Y}=\tan\varphi
\eea
we find that we recover Minkowski space,
\be
ds^2=dT^2-dX^2-dY^2-dZ^2\, .
\label{minko}\ee
Since
\be
X^2+Y^2+Z^2-T^2={1\over A^2}{y-x\over y+x},
\ee
it follows that the surface at $x=0$ where the domain wall lies,
corresponds in Minkowski space to the hyperboloid (see 
figure \ref{fig:wallinmink})
\be
X^2+Y^2+Z^2-T^2={1\over A^2}
\label{hype}
\ee
and the region $x>0$ is the interior of the hyperboloid
$X^2+Y^2+Z^2-T^2<1/A^2$. Sections at constant $T$ are spheres, so the
spatial geometry of the wall is spherical. In fact, the intrinsic
geometry of the wall is precisely that of de Sitter space in $2+1$
dimensions, as is evident from (\ref{desit}). 

Now we add the black holes, by allowing for $m>0$. In this case,
$y=+\infty$ is a singularity, surrounded by at least one black hole
horizon at $y=|\xi_2|$ (and an inner black hole horizon $y=|\xi_1|$ if
$q\neq 0$). This implies that, if the black hole is not too large, we
can get a good approximation for its position on the wall by looking at
the trajectories of points at large positive values of $y$ in the metric
(\ref{nobh}). In particular, the central singularity of the black hole,
at $y=+\infty$ is mapped to $Y^2+Z^2=0$, $X^2-T^2=1/A^2$, i.e.,
$X=\pm\sqrt{T^2+1/A^2}$, $Y=Z=0$. At any given instant $T$, these are
{\it two} points at antipodal points on the domain wall. Therefore, the
WBH metric actually describes two black holes at antipodal points of a
spherical domain wall. It is easy to see that the black holes must be
oppositely charged. 

\begin{figure}[ht]
\begin{center}\leavevmode  %
\epsfxsize=8cm 
\epsfbox{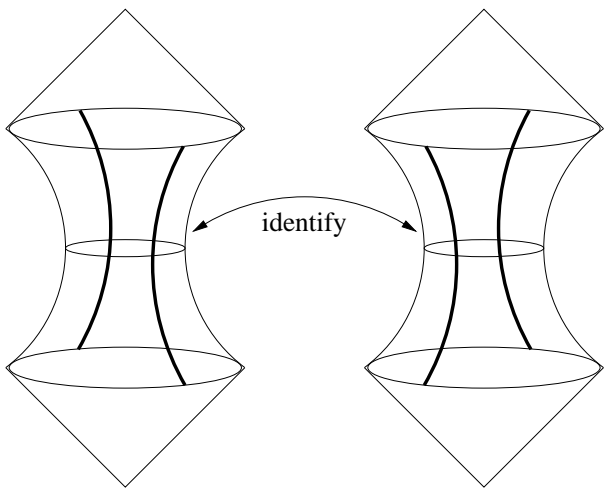}
\end{center}
\caption{Construction of the domain wall by gluing two copies of the
region $x>0$ along the surface $x=0$. The wall tension results from the
non-vanishing extrinsic curvature of this surface.} 
\label{fig:wallident}
\end{figure}

Note that the domain wall not only eliminates the string from
the spacetime, but it also cuts off the acceleration horizon and turns
it into a horizon of finite area. The entire construction is depicted in 
figure \ref{fig:wallident}.

Finally, another interesting point is that, as argued in
\cite{EHM3}, the black holes will neither swallow up the brane, nor
slide off of it. The latter was argued on the basis of the elastic
restoring force that the brane exerts on the black hole. Below we will
give an alternative, thermodynamic argument for this feature.

\subsection{Thermodynamics of a black hole on the wall}

The conventional definitions for the mass of the black hole, e.g.,
ADM, cannot be applied to a spacetime such as the present one, which is
not asymptotically flat. Nevertheless, black hole thermodynamics can be
used to give a prescription for the mass. This approach was put to use
in \cite{EHM}, where it was shown to yield very satisfactory results.
The method consists of identifying the black hole entropy using the area
formula $S={\cal A_{\rm bh}}/4G$, and the temperature $T$ in terms of
the surface gravity $\kappa$, $T=\kappa/2\pi$. Then, the first law of
thermodynamics (in the absence of charge),
\be
\delta M=T\delta S
\ee
can be integrated to give the black hole mass $M$. For
simplicity, we choose to set to zero the black hole charge in this
subsection. This means that $G(x)=1-x^2-2 mA x^3$, and
$\xi_1\to -\infty$, but we remain consistent with the notation so far,
i.e., $y=|\xi_2|$ corresponds to the location of the black hole horizon,
$0\leq x\leq \xi_4$, etc.\footnote{Notice that here
we are out of thermodynamic equilibrium, in the sense that the
temperature of the black hole and acceleration horizons are different.
This poses no problem at this point, but
see section \ref{sect:nbh} for more on this.} 

The calculation of the area is straightforward,
\bea\label{bharea}
{\cal A_{\rm bh}}&=&\int
dxd\varphi\sqrt{g_{xx}g_{\varphi\varphi}}|_{y=|\xi_2|}\nonumber\\
&=&2{\Delta\varphi\over A^2}\int^{\xi_4}_0 {dx\over (x+
|\xi_2|)^2}=2{\Delta\varphi\over A^2}{\xi_4\over |\xi_2|(|\xi_2|+
\xi_4)}\, .
\eea
The factor of $2$ in the second line arises from the
double-sided character of the wall.

The surface gravity is computed relative to the timelike Killing vector
$\partial_t$. In order for it to have the right dimensionality, we have
to multiply it by a factor of $A$, $\chi=A\partial_t$, and then
$\kappa^2=\chi_{\mu;\nu}\chi^{\mu;\nu}/2$. With this normalization for
$\chi$ one also recovers the standard result for the temperature of the
Schwarzschild-deSitter black hole on the brane (\ref{desit}). The result
is
\be
T={A|F'(-\xi_2)|\over 4\pi}= {A|G'(\xi_2)|\over 4\pi}\, .
\ee

In order to perform the integration of the first law it is convenient
to introduce the auxiliary variable
\be
z={|\xi_2|\over \xi_4}\, .
\ee
The limit where the black holes are absent, $m=0$, corresponds to
$z\to\infty$. On the other hand, there is a lower bound for $z$ imposed
by the fact that $G(x)$ must have three real roots. This requires
$mA<1/3\sqrt{3}$, hence $|\xi_2|>\sqrt{3}$ and $z>2$ (when this bound is
saturated the black hole and acceleration horizons coincide). Therefore
the range for $z$ is $2<z<\infty$.

In terms of $z$ we have
\bea
\xi_4&=&{\sqrt{z^2-z+1}\over z}\, ,\qquad
|\xi_2|=\sqrt{z^2-z+1}\, ,\\
mA&=&{z(z-1)\over 2(z^2-z+1)^{3/2}}\, ,\nonumber
\eea
and
\bea
S&=&{2\pi\over G A^2}{z\over (1+z)^2(2z-1)}\, ,\nonumber\\
T&=&{A\over 4\pi}{z^2-z-2\over \sqrt{z^2-z+1}}\, .
\eea
In the range $2<z<\infty$, $m$ and $S$ decrease monotonically with
$z$, while $T$ grows
monotonically with it.
The first law
\be
{d M\over dz}=T{d S\over dz}
\ee
can now be integrated, with the condition that $M=0$ for $m=0$,
i.e., for $z\to\infty$. 
In this way we find that the black hole mass is
\be
M={1\over G A}\left( {1\over 2}-{z\sqrt{z^2-z+1}\over 2z^2+z-
1}\right)\, .
\ee

\begin{figure}[ht]
\begin{center}\leavevmode  %
\epsfxsize=8cm 
\epsfbox{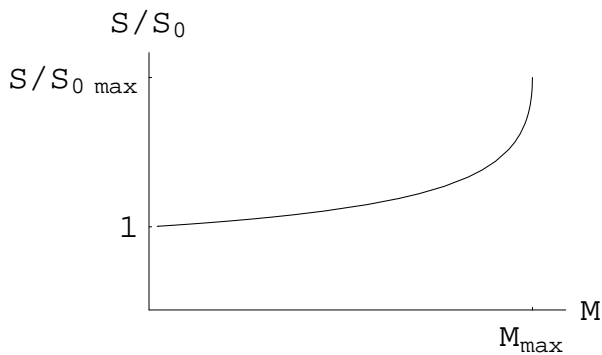}
\end{center}
\caption{The entropy of a black hole on the wall, $S$, compared to the
entropy of a black hole away from the wall, $S_0=4\pi GM^2$.
Notice $S>S_0$ for all allowed values of $M$. The maximum values
correspond to $z\to 2$.} 
\label{fig:entropy}
\end{figure}

Let us
examine what these formulas yield for small black holes, up to next
to leading order in $mA$. Using (\ref{approot}),
\bea
M&=&{m\over G}\left(1-{15\over 4}mA+O(m^2A^2)\right)\, ,\nonumber\\
S&=&{4\pi m^2\over G}\left(1-5 mA+O(m^2A^2)\right)\, .
\eea
If we now express the entropy in terms of the physical mass $M$,
and the wall tension $\sigma$,
\be
S={4\pi G M^2}\left(1+5\pi G^2\sigma
M+O(G^4 M^2\sigma^2)\right)\, .
\ee
The leading order result reproduces the standard formula for
Schwarzschild black holes. The first correction tells us that, for a
given black hole mass, the entropy will be higher if it is on a wall,
than if it is away from it. This feature persists throughout the entire
range of masses, as exhibited in figure \ref{fig:entropy}. Hence, it is
thermodynamically favored for a
black hole to stick to the wall, in accord with the above mentioned fact
that black holes do not slide off of the wall.

We end this analysis with two remarks related to the existence of an
upper limit for the size of the black holes on the wall. First, one may
notice that $z$ can be allowed a range wider than considered above,
namely, one can also have $1\leq z\leq 2$ without encountering any
singular behavior. However, while the limit $z\to 2$ from above
corresponds to $mA\to 1/3\sqrt{3}$ from below, going to $1\leq z<2$ does
not correspond to having larger $mA$ (since we still have three real
roots), but rather to exchanging $\xi_2$ and $\xi_3$, i.e., the black
hole and acceleration horizons. For $1\leq z\leq 2$, the plot in figure
\ref{fig:entropy} would extend to yield the entropy (area) of the
acceleration horizon, for a given black hole mass.

The second remark concerns an aspect of the shape of black holes on
branes discussed in ref.\cite{EHM}. There, it was argued that large
black holes in the four dimensional RS brane-world have the shape
of a pancake, the horizon being flattened and having a small extent away
from the brane. Is there anything similar in the present context? 
It would appear that there is not, at least not anything 
as drastic as in \cite{EHM}. The
crucial difference is that in the situation studied in \cite{EHM}, the
geometry of the brane was not asymptotically deSitter, but instead,
asymptotically flat. The ``black pancakes'' corresponded to $mA$
becoming larger than $1/3\sqrt{3}$. We have explicitly excluded this
from our analysis: the reason is that for $mA=1/3\sqrt{3}$ the black
hole horizon and the acceleration horizon (which yields the cosmological
deSitter horizon on the brane) become coincident, and for
$mA>1/3\sqrt{3}$ there is no black hole on the brane, but rather a naked
singularity. This obstruction obviously disappears for the
asymptotically flat brane of \cite{EHM}, and in that case the horizon on
the brane can grow arbitrarily large. There are no ``black
pancakes'' in our setting, and all our black holes are roughly
spherical. On the other hand, if there were a negative cosmological
constant in the bulk, but a positive cosmological constant induced on
the brane, the cosmological brane horizon would be larger than in the
present setting, and the upper limit on the size of the black holes
would be higher.

\section{Thin wall approximation}\label{sect:twa}

Having described in detail the wall-black hole spacetime in the limit of
infinitesimal wall thickness, we now proceed to demonstrate how the
distributional wall, with the black hole on it, can consistently arise
as a limit of a physical, field-theoretical topological defect.
This is needed to establish the analytic approximation which will be
used to derive the gravitational back reaction of the wall on the
spacetime. 

We will use a general field theory lagrangian for the wall
\be
8\pi G{\cal L}_{\rm DW} = {\epsilon \over w^2} \left [
w^2 (\nabla_a X)^2 - V(X) \right ],
\label{wallag}
\ee
where the symmetry breaking potential, $V$, has a discrete set of degenerate
minima, and we have rescaled the scalar field $X (= \Phi/\eta)$ by the
symmetry breaking parameter $\eta$. The parameter $\epsilon = 8\pi G\eta^2$ 
represents the gravitational strength of the domain wall, being the energy
density per unit area, and $w$ represents the inverse mass
of the scalar after symmetry breaking, which will also characterise the
width of the wall defect within the theory. Without loss of generality, we
will fix our units by setting $w=1$. The wall equations,
\be
\Box X + {1 \over 2} {\partial V \over \partial X} = 0,
\label{walleqs}
\ee
have the first integral $ X'^2 = V(X)$ in Minkowski spacetime,
which has an implicit solution
\be
\int^{X}_{X_{F}} {dX\over\sqrt{V(X)}} = z - z_0,
\label{flatx}
\ee
where $X_{F} = X(z_0)$ is the false vacuum. For example, in the
$\lambda\Phi^4$ kink model, $w = 1/\sqrt{\lambda}\eta=1$, and
the above integral (\ref{flatx}) gives the
usual kink solution centered on $z_0$: $X = \tanh(z-z_0)$,
which has an energy per unit area
\be
8\pi G \sigma = 2\epsilon \int_{-\infty}^\infty \sech^4z =
{8\epsilon\over3}. 
\label{epua}
\ee

Now let us consider the wall equations (\ref{walleqs}) in the 
Reissner-Nordstrom background
\be
ds^2 = \left (1 - {2M \over r} + {Q^2\over r^2} \right ) dt^2 - 
\left (1 - {2M \over r} + {Q^2\over r^2} \right ) ^{-1} dr^2 - r^2
d\theta^2 -
r^2 \sin^2\theta d\varphi^2,
\label{rnwall}
\ee
where $M$, and $Q$ are  measured in ``wall'' units, (i.e.\ $w=1$) rather
than Planck units, and the numerical value of $G$ in wall units has been
absorbed into $M$ and $Q$. This gives for the scalar $X$:
\be
\left (1 - {2M \over r} + {Q^2\over r^2} \right ) X_{,rr}
+ {2\over r} \left ( 1- {M\over r} \right ) X_{,r} +
{1\over r^2} X_{,\theta\theta} + {\cot\theta\over r^2} X_{,\theta}
= {1\over 2} {\partial V \over \partial X}\, .
\label{rneqs}
\ee

Recall that for the case of the vortex, \cite{AGK}, the fields were very
well approximated as functions of $r\sin\theta$, therefore, guessing
the Ansatz $X = X(\texttt{z})=X(r\cos\theta)$, we find that
\be
\Box X = - X''\left [ 1 - {2M\texttt{z}^2\over r^3} 
+{Q^2 \texttt{z}^2\over r^4} \right] + {2M\texttt{z}\over r^3} X'
\label{apprn}\, .
\ee
Now, noting that $r$ is strictly greater than $M$ outside the horizon of
an RN black hole, we see that the \texttt{z}-dependent terms in (\ref{apprn})
are of order \texttt{z}$M^{-2}$ or $\texttt{z}^2M^{-2}$. 
Therefore, if the thickness of the wall is much
less than the black hole horizon size, i.e.\ $M\gg1$, we see that
$X = X_0(\texttt{z})$, where $X_0$ is the flat space solution of
(\ref{flatx}), will solve the equations of motion up to O($M^{-2}$), since
the derivatives of $X_0$ differ from zero 
significantly only for \texttt{z} = O(1). Thus we see that a thin
wall can be painted on to a black hole solution, as confirmed by
the numerical work in \cite{MYIIN}.

However, since we expect our gravitating wall-black hole system to
have a metric of the form (\ref{Cmet}), we have to see how to paint 
the wall onto the C-metric, which at first sight is a rather 
different looking beast, however, if we change coordinates via
\be
{\bar t} = A^{-1}t \;\; , \;\; r = 1/Ay \;\; , \;\; {\rm and} \;\;\;
\theta = \int_x^{x_3} dx/\sqrt{G(x)}
\label{cmetcoords}
\ee
then
\be
ds^2 = [1 + Arx(\theta)]^{-2}
\left [ (1-{\textstyle{2m\over r}} + {\textstyle{q^2\over r^2}} 
- A^2r^2) d{\bar t}^2 
- {dr^2 \over (1-{2m\over r}  + {q^2\over r^2}- A^2r^2)} 
- r^2 d\theta^2 - r^2 G(x) d\varphi^2 \right ].
\ee
We see therefore that the variable $x$ is basically $\cos\theta$,
and therefore we guess that $\texttt{z} = x/Ay$. Substituting
this into the wave operator for $X$ gives
\bea
-\Box X &=& X'' \left ( 1 + A\texttt{z}\right)^2 \left [ G(x) + 
A^2 \texttt{z}^2 F(y) \right ]\\
&+& X' A \left ( 1 + A\texttt{z}\right)
\left [ 2 G(x) + 2 A \texttt{z} F(y) \left ( 2+ A\texttt{z}\right) +
\left ( y G'(x) - x F'(y) \right ) \left ( 1 + A\texttt{z}\right) 
\right ]\, .\nonumber
\eea
We now need to consider what the ``thin wall'' approximation means in the
context of the C-metric. Clearly we expect the black hole horizon radius
to be large in these units, i.e.\ $A|\xi_2|\ll1$, however, recall that
for a self-gravitating domain wall, there is a limit to wall formation
given by the size of the spontaneously compactified spacetime, which 
corresponds to the acceleration horizon. Therefore, although we are not
at this point considering gravitational back reaction, we will also 
work in the r\'egime of large acceleration radius, i.e.\ $A|\xi_3|\ll1$.
Now, the wall fields differ significantly from their vacuum values for
$\texttt{z} \sim 1$, and the values of $y$ are bounded by the black 
hole horizon, and the acceleration horizon $1<|\xi_3|\leq y < |\xi_2|$. 
Therefore $x\leq Ay \ll 1$ in the core of the wall. Meanwhile, the 
maximum value of $F(y)$ is at most of order $|\xi_2|^2$, hence we see that 
\be
-\Box X = X'' + O(A)\, .
\ee
So $X=X_0(\texttt{z})$ is indeed a good approximation to the solution 
of the field eqns in the C-metric background.

\section{Extremal horizons expel thick walls}\label{sect:expel}

Having argued the existence of the domain wall solution in the black hole
background for large mass black holes, we will now consider the special
case of an extremal black hole, for which the inner and outer horizons 
coincide. First consider the extremal RN black hole
\be
ds^2 = \left (1 - {M\over r} \right )^2 dt^2 - \left (1 - {M\over r} 
\right )^{-2} dr^2 - r^2 d\theta^2 - r^2 \sin^2 \theta d\varphi^2\, .
\ee
For the cosmic string, or a pure flux p-brane, a phenomenon of 
`flux expulsion' occurs for sufficiently small mass black holes,
namely, the defect ceases to penetrate the black hole horizon, and
instead wraps around it with the horizon remaining in the symmetric
phase across all of its area. If such a
phenomenon occurs with the domain wall, this would mean that
$X\equiv0$ over the event horizon, and all of the nontrivial field
dynamics of the wall would occur in the exterior region. Although the
above thin wall approximation indicates that for large mass black holes
there is a solution with the wall intersecting the black hole, it gives no 
indication of what might happen for $M\leq O(1)$. 

Indeed, a simple argument gives us a first indication that at least very
small extremal black holes sitting well inside the wall will expel it.
Deep inside the core of
the wall the potential terms are very small compared to the gradient
terms, so we neglect them. In this approximation, we now try to solve
the wall equations in the RN background, i.e., eq.(\ref{rneqs}) with
the RHS set to zero. Given that, in the absence of a black hole, the solution 
for a wall in the region close to its center is $X\approx 
\texttt{z}-\texttt{z}_0$, we try the
ansatz $X=b(r)\cos\theta$, and obtain the equation
\be
\left(r^2 -2Mr+Q^2\right)b''+2(r-M)b'-2b=0\, ,
\ee
which admits $b=r-M$ as the regular solution at the horizon, i.e.,
\be
X\approx (r-M)\cos\theta\, .
\ee
We see that if the black hole is not extremal, then $X\neq 0$ on the
horizon, but $X$ will vanish on an extremal horizon.
We now derive a more precise and rigorous bound on $M$ for which the
wall must be expelled from the black hole.

Suppose there is a solution to the equations of motion which penetrates the 
horizon, then on the horizon the field equations become
\be
X_{,\theta\theta} = - \cot\theta X_{,\theta} + 2M^2 X(X^2-1)\, ,
\ee
i.e.\ an ODE for $X$ in terms of $\theta$. Taking the derivative gives
\be
X_{,\theta\theta\theta} = - \cot\theta X_{,\theta\theta}
+ X_{,\theta} \left [ \csc^2\theta + 2M^2 (3X^2-1)\right]\, .
\label{xtripp}
\ee
Now, any nontrivial solution will satisfy $X({\pi\over2}) = 0$ which implies
that $X_{,\theta\theta}({\pi\over2}) = 0$. Without loss of generality we
may suppose that $X_{,\theta}({\pi\over2})>0$, so that $X_{,\theta}$
has a maximum (minimum) at $\pi\over2$ for $M^2>1/2$ ($M^2<1/2$) since
(\ref{xtripp}) implies that at any turning point of $X_{,\theta}$
\be
X_{,\theta\theta\theta} = X_{,\theta} \left [ 
\csc^2\theta + 2M^2 (3X^2-1)\right] > X_{,\theta}\left [1 - 2M^2\right]
\ee
(for $X_{,\theta}>0$). But we now see that if $M^2<1/2$, any turning point
of $X_{,\theta}$ is a minimum, which is inconsistent with
$X_{,\theta}=0$ at $\theta = 0,\pi$. Therefore for $M^2<1/2$, the only
possible solution is $X\equiv 0$ on the horizon, i.e.\ expulsion must
occur. By continuity, we in fact expect the true limit on $M^2$ for
expulsion to be somewhat higher than $1/2$, although numerical work
would be required to establish the true bound. For the C-metric, the
argument is slightly more involved, however, using the
$\theta$-coordinate as defined in (\ref{cmetcoords}) we can derive a
similarly weak bound for expulsion as $A|\xi_2|>1$. 

Now let us examine the region $M^2>1/2$. 
If a stable expelling solution exists, we expect that in the region
near the horizon, it monotonically relaxes to a kink solution as we 
move away from the horizon, namely, $X_{,\theta}$ remains of the 
same sign (without loss of generality we will take $X_{,\theta}>0$
for $r>M$). Let us consider the full field equations in a neighborhood 
of the horizon such that $r<M+\delta/M$ and $|X|<\delta$ 
for some small parameter $\delta$: 
\be
X_{,\theta\theta} = - \cot\theta X_{,\theta} - 2 M^2 X -
[(r-M)^2 X_{,r}]_{,r} + O(\delta)\, .
\label{xthth}
\ee
By considering $\theta$ derivatives of this equation, it is possible to
show that $X_{,\theta\theta},X_{,\theta\theta\theta}\leq0$ and
$X_{,\theta\theta\theta\theta}\geq0$ on $[{\pi\over2},\pi]$ 
for $M^2>1/2$, hence
\be
{\pi\over4}X_{,\theta}\left ({\pi\over2}\right ) < X(\pi)<
{\pi\over2}X_{,\theta}\left ({\pi\over2}\right )
\ee
and
\be
{\pi\over4}|X_{,\theta\theta}\left (\pi\right )| < X_{,\theta}({\pi\over2})<
{\pi\over2}|X_{,\theta\theta}\left (\pi\right )|\, .
\ee
Combining these inequalities, and reading off $X_{,\theta\theta}(\pi)$
from (\ref{xthth}) gives
\be
{\pi^2 M^2 \over 16} X(\pi) < X(\pi) < {\pi^2 M^2 \over 4} X(\pi)\, ,
\ee
i.e.\ $2/\pi < 1/\sqrt{2} < M < 4/\pi$. Therefore, if
$M > 4/\pi$ there can be no such expelling solution.

\section{Gravitational back reaction}\label{sect:gbr}

First let us briefly recall the self-gravitating domain wall. In wall-based
coordinates the metric can be written in the form
\be
ds^2 = a^2(z) \left [ dt^2 - e^{2kt} (dx^2+dy^2)\right ] - dz^2\, ,
\ee
where the function $A(z)$, and the constant $k$, can be determined
analytically as a power series in $\epsilon$, the leading order
values being
\bml\bea
a(z) &=& 1 - \epsilon \int_0^z dz' \int_0^{z'} dz'' V(X_0(z''))
= 1 - {\epsilon\over 6} \left [ 4\log\cosh z + \tanh^2z\right]\, ,\\
k &=& \epsilon \int_0^\infty dz' V(X_0(z')) = {2\epsilon\over 3}\, ,
\eea\eml
the explicit forms being those for the $\lambda \Phi^4$ kink.
Note that the change in extrinsic curvature from one side of
the kink to the other is $[a'] = -4\epsilon/3 = - 4\pi G\sigma$ from
(\ref{epua}) as required by the Israel junction conditions.
Comparing this with (\ref{extc}) for the WBH solution, we see that
we must have the acceleration parameter $A = 2\epsilon/3$ for 
consistency with the self-gravitating
thick wall.

In general, our thick-WBH metric will take the form
\be
ds^2 = \Omega^{-2} \left [ E^2 dt^2 - B^2 dy^2 - D^2 dx^2 - C^2
d\varphi^2 \right]\, ,
\label{canmet}
\ee
where to leading order $X=X_0$ and (\ref{canmet}) takes the form of 
(\ref{Cmet}). We will now perform a linearised calculation 
in $\epsilon = 3A/2$, the
energy of the wall, writing $\Omega = \Omega_0 + A \Omega_1$ etc.\ where
$\Omega_1/ \Omega_0 = O(1)$ near the core of the wall, and tends to
zero away from the core. 

Let's start by calculating the stress-energy of the $X_0$-field.
\bml\bea
g^{xx} X_{0,x} X_{0,x} &=& - G(x) A^2 (|x|+y)^2 \left ({X_0'\over Ay} 
\right )^2\simeq - X_0^{\prime2}\, , \\
g^{yy} X_{0,y} X_{0,y} &=& - F(y) A^2 (|x|+y)^2 \left
({-\texttt{z}X_0'\over y} 
\right )^2 = O\left ( (Ay_h)^2\right)\, ,
\eea\eml
Since the gauge potential for the Maxwell field is unaltered by the
presence of the wall, its energy momentum tensor remains formally the same:
\be
T^0_{_{\rm EM}0} = T^y_{_{\rm EM}y} = q^2 \Omega^4 = 
-T^x_{_{\rm EM}x} = -T^\varphi_{_{\rm EM}\varphi}\, .
\ee
Therefore the Einstein equations for the wall,
\be
R_{ab} = 2\epsilon X_{,a} X_{,b} - \epsilon V(X) g_{ab} + T_{_{\rm
EM}ab}\, ,
\ee
become to leading order in $A$ 
\be
R^0_0 - q^2 \Omega^4 = R^\varphi_\varphi + q^2\Omega^4 = -{3A\over2} V(X) 
= R^y_y - q^2 \Omega^4 + O(A^2)\; = {\third} ( R^x_x + q^2\Omega^4) +
O(A^2)\, .
\label{leadeeq}
\ee
Since in (\ref{Cmet}) the variation of the extrinsic curvature due to the
wall is primarily carried by $\Omega$, with $E_0^2 = B_0^{_2} = F(y)$ 
unaffected, and $C_0^2 = D_0^{-2} = G(x)$ only affected at $O(x^3)=O(A^3y^3)$
we guess that $E,B,C,D$ will effectively take their background values, 
in which case we find that the Ricci tensor is:
\bml\bea
R^0_0 &=& {\half} \Omega^2 F''(y) - 2F'(y) \Omega\Omega_{,y} - 
G'(x) \Omega\Omega_{,x} - \left[ G(x) \left ( 
\Omega\Omega_{,xx} - 3\Omega^2_{,x} \right)
+ F(y) \left ( \Omega\Omega_{,yy}- 
3\Omega^2_{,y} \right) \right]\\
R^\varphi_\varphi &=& {\half} \Omega^2 G''(x) - 2G'(x) \Omega\Omega_{,x} -
F'(y) \Omega\Omega_{,y}- \left[ G(x) \left (
\Omega\Omega_{,xx} - 3\Omega^2_{,x} \right)
+ F(y) \left ( \Omega\Omega_{,yy} - 
3\Omega^2_{,y} \right) \right]\\
R^x_x &=& {\half} \Omega^2 G''(x) - 2G'(x) \Omega\Omega_{,x} -
F'(y) \Omega\Omega_{,y}- \left[ 3G(x) \left (
\Omega\Omega_{,xx} - \Omega^2_{,x} \right)
+ F(y) \left ( \Omega\Omega_{,yy} - 
3\Omega^2_{,y} \right) \right]\\
R^y_y &=& {\half} \Omega^2 F''(y) - 2F'(y) \Omega\Omega_{,y} - 
G'(x) \Omega\Omega_{,x} - \left[ G(x) \left (
\Omega\Omega_{,xx} - 3\Omega^2_{,x} \right)
+ 3F(y) \left ( \Omega\Omega_{,yy} - 
\Omega^2_{,y} \right) \right]\\
R_{xy} &=& {\Omega_{,xy}\over\Omega}\, .
\eea\eml
The Einstein equations (\ref{leadeeq}) then suggest
\be
\Omega = A(f + y),
\ee
where $f_0 = |x|$. Inputting this Ansatz gives
\bml\bea
{\half} \left (R^\varphi_\varphi - R^x_x \right ) &=& 
G(x) A^2 y (1+A|\texttt{z}|) f_{,xx} = 
{3A\over 2} \sech^4{x\over Ay} \label{feqn}\\
{\half} \left (R^0_0-R^y_y\right ) &=&
F(y) A^2 y (1+A|\texttt{z}|) f_{,yy} = O(A^2)\\
{\half} \left (R^0_0 - R^\varphi_\varphi\right) - q^2 \Omega^4 &=& 
A^2(y+f) \Bigl[ (1+3mA|x|+2q^2A^2 x^2) (f-xf_{,x})  - yf_{,y} \hskip 1cm 
\nonumber\\
+ 3mAy(|x|&-&f+yf_{,y}) +q^2A^2 [ (x^2-f^2)(f+3y) - 2y^3f_{,y}] \Bigr] 
= O(A^2)\, .
\eea\eml
Using \cite{BCG}, (\ref{feqn}) has solution
\be
f = Ay \left [ \ln\cosh{x\over Ay} 
+ {1\over4} \tanh^2 {x\over Ay} \right]\, ;
\ee
note that while $f_{,x}=O(A)$, $f_{,y} = (f - xf_{,x})/y=O(A^2)$, and
the Einstein eqns are satisfied to leading order in $A$.

Therefore, the topological kink solution smooths out the shell-like
singularity of the infinitesimal domain wall in much the same way
as the topological local vortex solution smooths out the delta-function
singularity in the AFV and other metrics.

\section{Nucleation of black holes on walls}\label{sect:nbh}

Both cosmic strings and domain walls are objects with a tension that
tends to make them unstable to snapping or forming holes on them. In the
absence of gravity, they may be protected against such instabilities by
the topology of the field configuration that gives rise to them.
However, it has been known for some time now that even topologically
stable vortices are unstable to snapping by forming a pair of black
holes at their endpoints \cite{SMTH}. This is a quantum tunneling
process, mediated by an instanton obtained from the Euclidean
continuation of the same C-metric as in (\ref{Cmet}).

An investigation of a related instability for domain walls was initiated
in \cite{CCG}. It was found that domain walls could nucleate black holes
at a finite distance from them. That is, a spherical domain wall that is
accelerating in an otherwise empty Minkowski space (such as we
have described in eqs.(\ref{minko}-\ref{hype})) may tunnel to a
configuration where it encloses a black hole. Actually, given the
double-sided nature of the wall, it encloses two black holes, one on
each side of the wall. It should be noted that the description of such
tunneling process is not without its qualms, since given the compact
nature of the Euclidean solutions involved, the wall (and the entire
universe with it), must be annihilated before giving way to the
configuration where the wall encloses the black hole\footnote{This is
familiar also from processes of black hole nucleation in an inflating
universe \cite{MR}.}.

This instability of the domain wall, however, is not quite the
analogue of the snapping string. Instead, we shall describe now
how black holes can form {\it on} a domain wall (even on a topologically
stable one)---an instability that might be dubbed the {\it hole-punch}
mechanism.

The final state will be precisely the one we have been describing in the
previous sections: an accelerating spherical domain wall with a pair of
black holes grafted at antipodal points on the wall. The black holes do
not swallow the brane, hence the holes do not grow. Again, the
transition to this configuration involves a quantum fluctuation where
the initial domain wall is annihilated, to be recreated back with the
black holes on it. The probability for this to happen is given, in first
approximation, by $\exp[-(I-I_0)]$, where the Euclidean action $I_0$ of
the initial configuration (the wall without a black hole) is subtracted
from the Euclidean action $I$ of the final state with the black holes on
the wall (we are assuming no-boundary conditions for the wavefunctions
of the corresponding universes). Alternatively, $\exp[-(I-I_0)]$ can be
viewed as the ratio of the probabilities to nucleate a domain wall with
and without black holes riding on it.

Let us proceed to construct the wall-black-hole instanton.
After continuation $t\to i\tau$, in order to get the correct
signature we restrict the range of $y$ to $-\xi_3\leq
y\leq -\xi_2$. The endpoints of this interval are spherical bolts
(two-dimensional fixed point
sets of $\partial_\tau$) and in order to avoid the appearance of conical
singularities at them the Euclidean time coordinate has to be
periodically identified,
$\tau\sim\tau+\beta$, in such a way that
\be
\beta={4\pi\over G'(\xi_3)}={4\pi\over |G'(\xi_2)|}.
\ee
The second equality cannot be fulfilled unless we appropriately restrict
the parameters of the solution. Specifically, this equation requires
$\xi_1-\xi_2-\xi_3+\xi_4=0$.
The physical interpretation of this constraint is simple, and
corresponds to a condition of thermal equilibrium: the temperature of
the black hole must be the same as the one induced by the acceleration
horizon. A neutral black horizon is smaller, hence hotter, than an
acceleration horizon, and as a consequence the above equation has no
solutions if $q=0$. If charge is added to the black hole, its
temperature can be lowered and then tuned to match the acceleration
temperature. One could also consider extremal black holes, for which
$\xi_1=\xi_2$ so that the black hole horizon does not restrict $\beta$.
This can be accommodated easily in what follows, and does not lead to
qualitatively new results.

\begin{figure}
\begin{center}\leavevmode  %
\epsfxsize=5cm 
\epsfbox{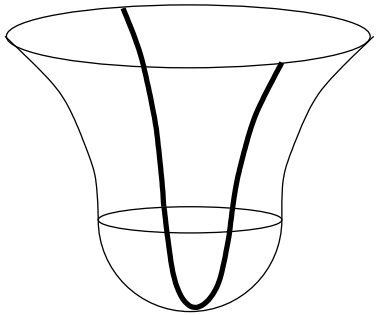}
\end{center}
\caption{The instanton process for nucleation of a domain wall with
black holes on it. The lower half-sphere is half the Euclidean
instanton, which then tunnels to real time and the hyperbolic expansion
of the wall. The black holes are created as a particle-antiparticle
pair. An identification of this geometry to a copy of itself along the
boundary $x=0$ must be performed, as in figure \ref{fig:wallident}.}
\label{fig:instanton}
\end{figure}

The hole-punch instanton thus constructed (see Fig.\ref{fig:instanton})
is completely non-singular
(save for the smoothable singularity at the location of the wall), and
it is also compact, which considerably facilitates the calculation of
its action. The latter can be computed directly by plugging the
explicit form of the solution into
\be
I=-{1\over 16\pi}\int d^4x \sqrt{g}(R-F^2)+\sigma\int d^3x\sqrt{h}\, .
\ee
($G=1$). This expression, in fact, can be further simplified on-shell by
use of the equations of motion which allow one to eliminate, say, $R$ in
favor of $F^2$ and the wall action. Alternatively, one can use the
on-shell equation \cite{HH}
\be
I=-{1\over 4}({\cal A}_{\rm acc}+{\cal A}_{\rm bh})\, ,
\ee
which gives the action in terms of the area of the horizons, which are all
finite in the present solution. Both methods yield, of course, the same
result, but the latter is perhaps slightly simpler.
The area of the black hole horizon has already been computed in
(\ref{bharea}), and that of the acceleration horizon is, analogously,
\be
{\cal A}_{\rm acc}=2{\Delta\varphi\over A^2}{\xi_4\over
\xi_3(\xi_3-\xi_4)}\, .
\ee

Collecting all terms, we obtain
\be
I=-{1\over 8\pi \sigma^2}{4\xi_4\over|G'(\xi_4)|}\left({1\over
\xi_3(\xi_3-\xi_4)}+{1\over\xi_2(\xi_2-\xi_4)}\right)\, .
\ee
On the other hand, the action for a domain wall instanton (without black
holes) is 
\be
I_0=-{1\over 8\pi\sigma^2}\, ,
\ee
and, as we said, the rate for nucleating holes on the wall is
$\sim\exp[-(I-I_0)]$. The resulting expression is not too illuminating,
but it simplifies in the limit of small black holes, $M,Q\ll
\sigma\propto 1/A$,
\be
I=-{1\over 8\pi \sigma^2}(1-8\pi M\sigma+O(M^2\sigma^2))
\ee
(recall that the black holes are nearly extremal, so $M\simeq Q$).
Hence,
\be 
I-I_0\simeq{M\over \sigma}\, ,
\label{actdiff}
\ee 
which, we observe, is a positive quantity. Nucleation of black holes is
exponentially suppressed, as was to be expected.  It
may be observed that, to this order, only ${\cal A}_{\rm acc}$
contributes. The black hole entropy $\sim 4\pi M^2$ contributes to
enhance the nucleation rate, but enters only at the next order.

We have demonstrated how, in addition to the process of nucleation of
black holes {\it enclosed by} a wall described in \cite{CCG}, black
holes can nucleate {\it on} the wall. The question arises of which
instability the domain wall is more likely to undergo. The two processes
are actually rather different, so one has to specify which
final states one is comparing. In \cite{CCG} it was found that the
action difference for 
nucleation of a neutral black hole inside the wall is
\be
I-I_0={11\sqrt{3}\over 36}{M\over \sigma}\approx 0.53{M\over\sigma},
\ee
which would be smaller than (\ref{actdiff}) and hence the process would
appear to be less suppressed. However, there is a crucial difference: in
the configuration where the black hole is enclosed by the wall, the mass
of the black hole is fixed to be
$M=(6\sqrt{3}\pi \sigma)^{-1}$ and hence cannot be varied independently
of the tension of the wall. In fact, the geometry is that of the
Schwarzschild solution, with the wall sitting at a fixed radius $r=3M$.
In other words, domain walls can only nucleate inside them black holes
of a certain (large) size. As a result, the process of \cite{CCG} can
only lead to the formation of very large black holes, and therefore will be
heavily suppressed. In contrast, in the hole-punch process the black
hole mass $M$ is a parameter that can be varied independently of
$\sigma$. Therefore, (\ref{actdiff}) can be made arbitrarily
small\footnote{The semiclassical approximation, however, will break down
when the black hole mass reaches the Planck scale.}. We
conclude that domain walls will preferentially
nucleate small black holes on them, rather than large ones inside them.

One might still be interested in comparing the processes where the
domain walls nucleate black holes with the same features---we take
the black holes to be extremal or nearly extremal. To this effect, we
should compare (\ref{actdiff}) to the nucleation rate of a (nearly)
extremal charged black hole inside the wall. In that case, $M$ is again
fixed in terms of $\sigma$, $M=Q=1/(8\pi \sigma)$, and \footnote{Of all
the situations considered in \cite{CCG} we are only taking the process
for which the action is smaller.}
\be
I-I_0={3\over 32\pi \sigma^2}={3\over 4}{M\over \sigma}\, ,
\label{actdiff3}
\ee
again smaller than (\ref{actdiff}). However, this result is exact,
whereas corrections to (\ref{actdiff}) (which can be seen to lower its
value) will be quite important since if we are supposed to be taking
$mA\simeq 2\pi M\sigma=1/4$, not a small number. And, in any event, the 
hole-punch process can form nearly extremal black holes much smaller
than this, which will be less suppressed.

\section{Conclusions}\label{sect:concl}

In this paper we have considered the problem of having a black hole
sitting on a topological domain wall, necessarily including the
gravitational back reaction. A domain wall has a very strong effect on
the spacetime surrounding it, causing a compactification of spatial
sections. We started by deriving the metric for an infinitesimally thin
domain wall bisecting a black hole, using the C-metric in a
recently-developed construction \cite{EHM}. The global structure of this
spacetime is the interior of two hyperboloids in a Lorentzian spacetime
(see figure \ref{fig:wallident}) with {\it two} accelerating black holes
`glued' to these walls. If the horizons are identified, then the black
holes are joined by a wormhole---this indeed happens if the black holes
are nucleated as a pair. We have used thermodynamics to provide a
definition for the mass of the black hole, which led us to
conclude that it is entropically preferred to have the black hole on the
wall, rather than away from it. We showed how one can smooth out the
`singular' behaviour of the zero thickness wall by using a core of a
topological, and hence thick, domain wall. Meanwhile for extreme black
holes, while the picture is qualitatively the same, if the wall is thick
enough relative to the black hole (roughly bigger than the black hole
size) the black hole will expel its flux, in the sense that the scalar
field forming the wall will remain in its false vacuum,
restored-symmetry state on the event horizon. This phenomenon might have
consequences for any brane-world model in which such charged black holes
are possible---recall that these black holes are charged under a gauge
field that is {\it not} confined to the brane. If one tries to pull a
small extremal black hole out of the brane, it will not experience the
same elastic restoring force that non-extremal black holes suffer, so,
apparently they might be able to slip off of the brane into the bulk. 

The problem of black hole nucleation on the domain wall -- the hole-punch
process -- is also analysed and we conclude it is more probable than the
black hole nucleation away from the wall considered in \cite{CCG}.

We want to conclude by discussing the possible extension of the results
in this paper to one dimension higher, this is, to the scenario with a
three-brane in five dimensions, which is pertinent for brane world
models. 

The first, and very serious, obstacle is the lack of an analogue
of the C-metric solution in five dimensions. Even if it is fairly safe
to assume, on physical grounds, that such a spacetime must exist, no
explicit construction of it has been found. Hence, the five dimensional
analogue of the wall-black hole metric remains unknown. In contrast,
there should be no problem in studying a black hole intersected by a
domain wall in five dimensions if the gravitational backreaction of the
latter is ignored. Including the backreaction in a perturbative fashion
might give some clues to the full solution. On
the other hand, the phenomenon of flux expulsion appears to be
essentially dimension independent.

The process of nucleation of black holes on walls presents some
peculiarities in five dimensions. Again, the absence of an explicit
analogue of the C-metric precludes a conclusive analysis. Nevertheless,
in \cite{GS} an instanton was presented which mediated the nucleation of
a domain wall in five dimensions, with a ``black object" on it. On the
brane, as well as on any surface at constant radius away from the brane,
the four-geometry is that of the Nariai instanton $S^2\times S^2$ (a
limiting case of the Euclidean Schwarzschild-deSitter solution). Hence,
the black object does not seem to be a black hole localized on the
brane, but rather a black string extending throughout the bulk---note,
however, that the authors of \cite{GS} argue otherwise. The size of the
black objects nucleated via this instanton is, as was the case in
\cite{CCG}, fixed by the wall tension. This is, the mass of the black
object is {\it not}\/ an independent parameter. In contrast, we have
argued that, in the four dimensional setting, a domain wall can nucleate
black holes of arbitrarily small size---which are preferred over larger
black holes. It seems reasonable to assume that a similar process will
be possible in five dimensions, and small black holes will be nucleated
on the brane. Nevertheless, notice that Euclidean regularity demanded
that the black hole be endowed with charge with respect to a bulk gauge
field. Such gauge fields are not always present in brane world models.
The absence of an explicit wall-black hole solution in five dimensions
leaves the door open to unexplored alternatives, but it might be that,
if no such charges are allowed, then the only possible instanton for
nucleation of walls with black objects on them, were that of \cite{GS}.


\section*{Acknowledgements}

We would like to thank Christos Charmousis for discussions. This work
has been supported by a Royal Society European Exchange Program grant.
RE acknowledges partial support from UPV grant 063.310-EB187/98 and
CICYT AEN 99-0315, RG is supported by the Royal Society, and CS is
supported by Funda\c c\~ao Ciencia e Tecnologia BPD/22092/9.

\end{document}